\documentclass[onecolumn,aps,amsmath,amssymb,superscriptaddress,pra]{revtex4-1}   
\usepackage[latin2]{inputenc}
\usepackage{amsmath,amssymb}
\usepackage{amsfonts}
\usepackage{bm}
\usepackage{dcolumn}
\usepackage{setspace}
\usepackage{graphicx}
\usepackage{multirow}
\usepackage{verbatim}
\usepackage{epstopdf}
\usepackage[normalem]{ulem}

\def\bar{\begin{array}}
\def\ear{\end{array}}

\def\f{\frac}

\def\nn{\nonumber}

\def\p{\partial}
\def\I{\mathcal{I}}
\def\Re{\mathrm{Re}}
\def\Im{\mathrm{Im}}
\def\ot{\theta}
\def\ox{\overline{x}}

\begin{document}

\title{Geometric energy transfer in two-component systems}

\author{Ryan Requist} 
\affiliation{
Fritz Haber Center for Molecular Dynamics, Institute of Chemistry, The Hebrew University of Jerusalem, Jerusalem 91904, Israel
}
\author{Chen Li} 
\affiliation{
Beijing National Laboratory for Molecular Sciences, College of Chemistry and Molecular Engineering, Peking University, Beijing 100871, China
}
\author{E. K. U. Gross}
\affiliation{
Fritz Haber Center for Molecular Dynamics, Institute of Chemistry, The Hebrew University of Jerusalem, Jerusalem 91904, Israel
}
\email{ryan.requist@gmail.com}

\date{\today}

\begin{abstract}
Factoring a wave function into marginal and conditional factors partitions the subsystem kinetic energy into two terms.  The first depends solely on the marginal wave function, through its gauge-covariant derivative, while the second depends on the quantum metric of the conditional wave function over the manifold of marginal variables.  We derive an identity for the rate of change of the second term. 
\end{abstract}

\maketitle

\section{Introduction}

Exposing a molecule to a visible or UV electromagnetic pulse excites primarily the electronic degrees of freedom.  On a longer time scale, the electrons transfer energy to the nuclei.  
While energy transfer can be evaluated with knowledge of the electron-nuclear wave function, given a choice of nuclear subsystem, to gain deeper insight into excited state molecular dynamics, optimize control of molecular motion and bond breaking, and engineer molecular scale mechanical motion, it would be valuable to have a formula expressing the energy transfer in terms of simple fundamental quantities.

Using the exact factorization (EF) formalism \cite{gidopoulos2014,abedi2010}, which 
involves the factorization of a two-component wave function into a product of marginal and conditional amplitudes \cite{hunter1975}, we have derived Ehrenfest identities \cite{li2019arxiv} for the time rate of change of the kinetic energy, momentum and angular momentum of the nuclear subsystem.  By evaluating the commutators in the Heisenberg equations of motion, we found that a novel quantum mechanical force operator appeared in all three identities.  In terms of this force operator, all three identities could be put in a form directly analogous to the classical equations of motion $dE/dt = \mathbf{F}\cdot \mathbf{v}$, $d\mathbf{P}/dt = \mathbf{F}$ and $d\mathbf{L}/dt = \mathbf{R} \times \mathbf{F}$.

One wrinkle in those results was the fact that the classical force-times-velocity form could only be achieved for the rate of change of one part of the nuclear kinetic energy, namely the term
\begin{align}
T_{n,marg} = \big< \chi \big| \sum_{i=1}^{N_n} \f{1}{2M_i} (\hat{\mathbf{P}}_{i} + \hat{\mathbf{A}}_{i})^2 \big| \chi \big> {,}
\end{align}
where $|\chi\rangle$ is the nuclear wave function that satisfies the effective Schr\"odinger equation
\begin{align}
i\p_t \chi(R,t) = \sum_{i=1}^{N_n} \f{1}{2M_i} \big[-i\nabla_{\mathbf{R}_{i}} + \mathbf{A}_{i}(R,t)\big]^2 \chi(R,t) + \mathcal{E}(R,t) \chi(R,t)\label{eq:schroedinger:nuclear}
\end{align}
of the exact factorization method; $R$ denotes the set $\{\mathbf{R}_i\}$ of $N_n$ nuclear coordinates.  $T_{n,marg}$ is the kinetic energy of a fictitious closed quantum system acted upon by scalar and vector potentials $\mathcal{E}(R,t)$ and $\mathbf{A}_{i}(R,t)$ (defined below).  In other words, it is simply the kinetic energy one would write down if one were given the Schr\"odinger equation (\ref{eq:schroedinger:nuclear}) but not told it describes only one component of a two-component system.  Since $\chi$ is the marginal probability amplitude for the nuclear component of a system of electrons and nuclei, we call $T_{n,marg}$ the marginal nuclear kinetic energy. 

The closed system described by equation (\ref{eq:schroedinger:nuclear}) reproduces the true $N_n$-body 
density and $N_n$-body current density of the nuclear subsystem, 
but it is fictitious in the sense that off-diagonal elements of the 
nuclear density matrix and hence nonlocal observables of the nuclear 
subsystem cannot be calculated directly from the wave function $\chi$. 
A prime example is the true nuclear kinetic energy as obtained from the 
full wave function $|\Psi\rangle$ which obeys the identity \cite{abedi2012,agostini2015}
\begin{align}
T_n = \big< \Psi \big| \sum_{i=1}^{N_n} \f{\hat{\mathbf{P}}_{i}^2}{2M_i} \big| \Psi \big> = T_{n,marg} + T_{n,geo} {,} \label{eq:Tn}
\end{align}
where the additional contribution is 
\begin{align}
T_{n,geo} = \int dR |\chi(R)|^2  \sum_{i=1}^{N_n} \f{1}{2M_i} \big( \langle \nabla_{\mathbf{R}_{i}} \Phi | \nabla_{\mathbf{R}_{i}} \Phi \rangle - |\mathbf{A}_{i}|^2 \big)  \label{eq:Tngeo:1}
\end{align}
with $\mathbf{A}_i=\Re\langle \Phi | \mathbf{P}_i \Phi \rangle$ and $\Phi(r|R) = \Psi(r,R)/\chi(R)$ being the electronic wavefunction conditional on $R$.  The Ehrenfest identity of Ref.~\cite{li2019arxiv} provides a straightforward way to calculate $dT_{n,marg}/dt$ but gives no information about $dT_{n,geo}/dt$.

Our main result is an exact identity for $dT_{n,geo}/dt$, expressed in terms of primitive quantities.  An interesting outcome of the derivation is the appearance of a new rank-3 quantum geometric quantity that cannot be expressed in terms of lower-rank quantities.

The integrand of (\ref{eq:Tngeo:1}) has geometric significance itself, as it can be expressed \cite{requist2016a,requist2016b} in terms of a tensor contraction between $\I^{\mu\nu}$, a symmetric inverse inertia tensor, and a Riemannian metric \cite{provost1980}
\begin{align}
g_{\mu\nu} = \langle (P_{\mu}-A_{\mu}) \Phi | (P_{\nu}-A_{\nu}) \Phi \rangle  \label{eq:g}
\end{align}
on nuclear configuration space, i.e.~the manifold $\mathcal{Q}$ with generalized coordinates $\{Q^{\mu}\}$ collectively denoted $Q$.  Thus, we have 
\begin{align}
T_{n,geo} = \int dR |\chi(R)|^2 \f{1}{2} \I^{\mu\nu} g_{\mu\nu} {,} \label{eq:Tngeo:2}
\end{align}
which we call the geometric part of the nuclear kinetic energy.  As all quantities except $\I^{\mu\nu}$ are time dependent, we suppress the time argument here and hereafter.  In (\ref{eq:g}) and (\ref{eq:Tngeo:2}), we have switched to a tensor calculus notation, i.e.~subscripts/superscripts correspond to the covariant/contravariant indices of a tensor on $\mathcal{Q}$, and we have generalized to a Watsonian kinetic energy operator $\hat{T}_n = (1/2) \I^{\mu\nu} P_{\mu} P_{\nu}$ \cite{watson1968} with $P_{\mu} = -i\p/\p Q^{\mu}$ and an implicit summation convention. The metric in equation (\ref{eq:g}) is the EF counterpart of the quantum metric tensor originally studied in the Born-Oppenheimer (BO) approximation \cite{berry1989,shapere1989,berry1990,berry1993}.  The quantum metric tensor has recently attracted attention in condensed matter physics, where its applications are too numerous to cite here.

The usual laboratory frame kinetic energy operator, cf.~equation (\ref{eq:Tn}), is a special case of the Watsonian kinetic energy, in which the set of $Q^{\mu}$ is $\{R_{1x},R_{1y},R_{1z},R_{2x},R_{2y},R_{2z},\ldots\}$ and $\I^{\mu\nu}$ is diagonal and $Q^{\mu}$-independent, i.e.~$\I^{\mu\nu}=\I^{i\alpha,j\beta}=M_i^{-1} \delta_{ij} \delta_{\alpha\beta}$ with $i$ labeling the nucleus and $\alpha=x,y,z$. 
The Watson form encompasses two additional cases: (1) an isolated molecule after removing the center-of-mass and overall rotational coordinates \cite{sutcliffe2000}; and (2) a system described by a distinguished set of relevant collective coordinates. The phonon modes of a crystalline solid described with Born-von Karman boundary conditions are an example of case (2) in which $Q^{\mu}$ comprise the normal mode amplitudes $U_{\mathbf{q}\lambda}$ with quasimomentum $\mathbf{q}$ and branch $\lambda$, $\I^{\mu\nu}=\I^{\mathbf{q}\lambda,\mathbf{q}'\lambda'}=\mathcal{M}_{\mathbf{q}\lambda}^{-1} \delta_{\mathbf{q},-\mathbf{q}} \delta_{\lambda,\lambda'}$ is off-diagonal and $\mathcal{M}_{\mathbf{q}\lambda}$ is the effective mass of the normal mode \cite{requist2019}.  In case (2), $\I^{\mu\nu}$ will be $Q^{\mu}$-dependent when $\mathcal{Q}$ is non-Euclidean, which can arise e.g.~in an approximate reduced description in terms of a restricted set of collective coordinates. In the case of an isolated molecule in which only the center-of-mass coordinate is removed, it is always possible to choose translationally-invariant coordinates such that $\I^{\mu\nu}$ is $Q^{\mu}$-independent \cite{requist2016b,sutcliffe2000}; such an $\I^{\mu\nu}$ is generally non-diagonal but can be further diagonalized.  Here, for simplicity, we assume $\I^{\mu\nu}$ is $Q^{\mu}$-independent but not necessarily diagonal.

\section{Exact factorization formalism}

To derive an identity for $dT_{n,geo}/dt$ that does not invoke the BO approximation, we make use of the exact factorization formalism \cite{hunter1975,gidopoulos2014,abedi2010}, the essential elements of which we briefly review here.

Starting from the full wavefunction $\Psi(q,Q) = \langle q,Q|\Psi\rangle$, where $q$ and $Q$ denote the sets of electronic and nuclear coordinates, one defines the nuclear wavefunction 
\begin{align}
\chi(Q) = e^{i\lambda(Q)} |\chi(Q)| {,}
\end{align}
which is the marginal probability amplitude corresponding to the marginal probability
\begin{align}
|\chi(Q)|^2 = \int dq |\Psi(q,Q)|^2
\end{align}
and the arbitrary gauge $\lambda(Q)$.  The conditional electronic wavefunction
\begin{align}
\Phi(q|Q) = \f{\Psi(q,Q)}{\chi(Q)}
\end{align}
depends parametrically on $Q$ and satisfies the equation 
\begin{align}
i\p_t |\Phi\rangle = (\hat{H}^{BO}-\mathcal{E}) |\Phi\rangle + \f{1}{2} (P_{\mu}-A_{\mu}) \I^{\mu\nu} (P_{\nu}-A_{\nu}) |\Phi\rangle + \f{(P_{\mu}+A_{\mu})\chi}{\chi} \I^{\mu\nu} (P_{\nu}-A_{\nu}) |\Phi\rangle {,} \label{eq:schroedinger:elec}
\end{align}
where $\hat{H}^{BO} = \hat{H} - \hat{T}_n$ and $\mathcal{E}(Q) = \langle \Phi(Q) | \hat{H}^{BO} | \Phi(Q) \rangle + \mathcal{E}_{geo}(Q)$ with $\mathcal{E}_{geo}(Q) = (1/2) \I^{\mu\nu} g_{\mu\nu}$; $\hat{H}$ is the usual nonrelativistic molecular Hamiltonian in atomic units.
Time and other arguments of functions will often be suppressed.   

\section{Derivation of the main result}

Starting from equation (\ref{eq:Tngeo:2}), we obtain
\begin{align}
\f{dT_{n,geo}}{dt} &= \int dQ \f{\p|\chi(Q)|^2}{\p t} \mathcal{E}_{geo}(Q) + \int dQ |\chi(Q)|^2 \p_t \mathcal{E}_{geo}(Q) \nn \\
&= \int dQ [ -\p_{\mu} J^{\mu}(Q) ] \mathcal{E}_{geo}(Q) + \int dQ |\chi(Q)|^2 \p_t \mathcal{E}_{geo}(Q) \nn \\
&=\int dQ J^{\mu}(Q) \p_{\mu} \mathcal{E}_{geo}(Q)  + \int dQ |\chi(Q)|^2 \p_t \mathcal{E}_{geo}(Q) {,}
\label{eq:dTngeo:1}
\end{align}
where $J^{\mu}$ is the nuclear current density  
\begin{align}
J^{\mu} = \I^{\mu\nu} \mathrm{Re} \big[ \chi^* (P_{\nu}+A_{\nu}) \chi \big]{.}
\end{align}
In the second line of equation (\ref{eq:dTngeo:1}), we have used the continuity equation, and in the third line, we have performed integration by parts and assumed that the boundary term vanishes; this is generally true for finite systems. Focusing on $\mathcal{E}_{geo} = (1/2) \I^{\mu\nu} g_{\mu\nu}$ in the second term and noting that $\I^{\mu\nu}$ is time-independent, our first step is to evaluate $\p g_{\mu\nu}/\p t$:
\begin{align}
\f{\p g_{\mu\nu}}{\p t} 
&= \Re \big< (P_{\mu}-A_{\mu}) \Phi \big| (P_{\nu}-A_{\nu}) \partial_t \Phi \big> 
+ \Re \big< (P_{\mu}-A_{\mu}) \partial_t \Phi \big| (P_{\nu}-A_{\nu}) \Phi \big>
{.} \label{eq:dgdt}
\end{align}
After contraction with $\I^{\mu\nu}$, these two terms give equal contributions to $\p_t \mathcal{E}_{geo}$.  There are no $\p_t A_{\mu}$ contributions because $\Re \langle \Phi |(P_{\mu}-A_{\mu}) \Phi \rangle = 0$.  The terms on the right-hand side of equation (\ref{eq:schroedinger:elec}) give the following contributions to the first term in equation~(\ref{eq:dgdt}):
\begin{align}\label{eq:dgdt:1abc}
\begin{split}
\f{\p g_{\mu\nu}^{(1a)}}{\p t} 
&= -\Re \big< (P_{\mu}-A_{\mu})\Phi \big| \partial_{\nu} \hat{H}^{BO} \big| \Phi \big> +\Im \langle (P_{\mu}-A_{\mu})\Phi \big| (\hat{H}^{BO} - \mathcal{E}) | (P_{\nu}-A_{\nu})\Phi \rangle \\
\f{\p g_{\mu\nu}^{(1b)}}{\p t} &= \Im \big< (P_{\mu}-A_{\mu}) \Phi \big| (P_{\nu}-A_{\nu}) \f{1}{2} \I^{\sigma\tau} (P_{\sigma}-A_{\sigma}) (P_{\tau}-A_{\tau}) \Phi \big> \\
\f{\p g_{\mu\nu}^{(1c)}}{\p t} &= \Im \big< (P_{\mu}-A_{\mu}) \Phi \big| (P_{\nu}-A_{\nu}) \I^{\sigma\tau} \f{(P_{\sigma}+A_{\sigma}) \chi}{\chi} (P_{\tau}-A_{\tau}) \Phi \big> {.}
\end{split}
\end{align}
The (1a) term can be put in various forms but none is simpler than the others.  The second term on the right-hand side of $\p g_{\mu\nu}^{(1a)}/\p t$ vanishes upon contraction with the symmetric tensor $\I^{\mu\nu}$ because 
\begin{align}
\I^{\mu\nu} \big< (P_{\mu}-A_{\mu}) \Phi \big| (\hat{H}^{BO} - \mathcal{E}) \big| (P_{\nu}-A_{\nu}) \Phi \big>
\end{align}
is real.  Turning to the (1b) term and moving $(P_{\nu}-A_{\nu})$ into the bra, we obtain
\begin{align}
\f{\p g_{\mu\nu}^{(1b)}}{\p t} 
&= - \f{1}{2} \I^{\sigma\tau} \p_{\nu} \Re \big< (P_{\mu}-A_{\mu}) \Phi \big| (P_{\sigma}-A_{\sigma}) (P_{\tau}-A_{\tau}) \Phi \big> \nn \\
&\quad + \f{1}{2} \I^{\sigma\tau} \Im \big< (P_{\nu}-A_{\nu}) (P_{\mu}-A_{\mu}) \Phi \big| (P_{\sigma}-A_{\sigma}) (P_{\tau}-A_{\tau}) \Phi \big> {.}
\label{eq:dgdt:1b}
\end{align}
The second term drops out upon contraction with the symmetric tensor $\I^{\mu\nu}$ because 
\begin{align}
\big< \I^{\mu\nu} (P_{\nu}-A_{\nu}) (P_{\mu}-A_{\mu}) \Phi \big| \I^{\sigma\tau} (P_{\sigma}-A_{\sigma}) (P_{\tau}-A_{\tau}) \Phi \big>
\end{align}
is real.  
Therefore, the contribution of the (1b) term to $\p \mathcal{E}_{geo}/\p t$ is
\begin{align}
\f{\p \mathcal{E}_{geo}^{(1b)}}{\p t} &= -\f{1}{4} \I^{\mu\nu} \p_{\nu} C_{\mu\sigma\tau} \I^{\sigma\tau} {,}
\label{eq:dEgeo:1b}
\end{align}
where we defined 
\begin{align}
C_{\mu\nu\tau} &= \Re \big< (P_{\mu}-A_{\mu}) \Phi \big| (P_{\nu}-A_{\nu}) (P_{\tau}-A_{\tau}) \Phi \big> {,}
\label{eq:C}
\end{align}
a manifestly gauge invariant quantity that will be examined in section~\ref{sec:CD}.  

For the (1c) term, we obtain
\begin{align}
\f{\p g_{\mu\nu}^{(1c)}}{\p t} 
&= \Im \bigg[ \big< (P_{\mu}-A_{\mu}) \Phi \big| (P_{\nu}-A_{\nu}) (P_{\tau}-A_{\tau}) \Phi \big> \I^{\tau\sigma} \f{(P_{\sigma}+A_{\sigma})\chi}{\chi} \bigg] \nn \\
&+ \Im \bigg[ \big< (P_{\mu}-A_{\mu}) \Phi \big| (P_{\tau}-A_{\tau}) \Phi \big> \I^{\tau\sigma}P_{\nu} \f{(P_{\sigma}+A_{\sigma})\chi}{\chi} \bigg] {.} 
\label{eq:dgdt:1c} 
\end{align}
The first term simplifies to
\begin{align}
\f{\p g_{\mu\nu}^{(1c.i)}}{\p t} 
&=  -\f{1}{2} C_{\mu\nu\tau} \I^{\tau\sigma} \f{\partial_{\sigma} |\chi|^2}{|\chi|^2} + D_{\mu\nu\tau} \f{J^{\tau}}{|\chi|^2} {,}
\label{eq:dgdt:1c.i}
\end{align}
where we introduced the definition
\begin{align}
D_{\mu\nu\tau} &= \Im \big< (P_{\mu}-A_{\mu}) \Phi \big| (P_{\nu}-A_{\nu}) (P_{\tau}-A_{\tau}) \Phi \big> {.} \label{eq:D}
\end{align}
The second term of equation (\ref{eq:dgdt:1c}) gives
\begin{align}
\f{\p g_{\mu\nu}^{(1c.ii)}}{\p t} &= -g_{\mu\tau} \p_{\nu} \f{J^{\tau}}{|\chi|^2} - \f{1}{4} B_{\mu\tau} \I^{\tau\sigma} \p_{\nu} \partial_{\sigma} \ln |\chi|^2 {,} 
\label{eq:dgdt:1c.ii}
\end{align}
where $B_{\mu\nu}=\p_{\mu} A_{\nu} - \p_{\nu} A_{\mu}$ is the Berry curvature.
Summing equations (\ref{eq:dgdt:1c.i}) and (\ref{eq:dgdt:1c.ii}) and contracting with $\f{1}{2}\I^{\mu\nu}$, we obtain  
\begin{align}
\f{\p\mathcal{E}_{geo}^{(1c)}}{\p t} &= -\f{1}{4} \I^{\mu\nu} C_{\mu\nu\tau} \I^{\tau\sigma} \f{\partial_{\sigma} |\chi|^2}{|\chi|^2} -\f{1}{2} \f{J^{\tau}}{|\chi|^2} \partial_{\tau} \mathcal{E}_{geo} - \f{1}{2} \I^{\mu\nu} g_{\mu\tau} \p_{\nu} \f{J^{\tau}}{|\chi|^2} {,}
\label{eq:dEgeo:1c}
\end{align}
where we used equation (\ref{eq:D:2}), derived in the section~\ref{sec:CD}, to relate $D_{\mu\nu\tau}$ to derivatives of $g_{\mu\nu}$, and hence to $\p_{\tau} \mathcal{E}_{geo}$.  The antisymmetry of $B_{\mu\nu}$ makes the contribution from the second term of equation (\ref{eq:dgdt:1c.ii}) vanish.  The second term of equation (\ref{eq:dEgeo:1c}) will be seen to cancel with the first term of equation (\ref{eq:dTngeo:1}).

Putting equations~(\ref{eq:dgdt:1abc}), (\ref{eq:dEgeo:1b}) and (\ref{eq:dEgeo:1c}) together, we obtain 
\begin{align}
\f{\p \mathcal{E}_{geo}}{\p t} &= - \I^{\mu\nu} \Re \langle \Phi | \partial_{\mu} \hat{H}^{BO} | (P_{\nu}-A_{\nu}) \Phi \rangle - \f{1}{2} \I^{\mu\nu} \partial_{\nu} C_{\mu\sigma\tau} \I^{\sigma\tau} 
- \f{1}{2} \I^{\mu\nu} C_{\mu\nu\tau} \I^{\tau\sigma} \f{\partial_{\sigma} |\chi|^2}{|\chi|^2} \nn \\
&\quad -\f{J^{\tau}}{|\chi|^2} \partial_{\tau} \mathcal{E}_{geo} - \I^{\mu\nu} g_{\mu\tau} \p_{\nu} \f{J^{\tau}}{|\chi|^2} {.}
\label{eq:dEgeo:final}
\end{align}
Substituting this into equation~(\ref{eq:dTngeo:1}) yields the final result
\begin{align}
\f{dT_{n,geo}}{dt} 
&= -\int dQ |\chi|^2 \I^{\mu\nu} \Re \langle \Phi | \partial_{\mu} \hat{H}^{BO} | (P_{\nu}-A_{\nu}) \Phi \rangle - \int dQ \f{1}{4} \I^{\tau\sigma} \p_{\sigma} B_{\tau\mu} \I^{\mu\nu} \partial_{\nu} |\chi|^2 
 \nn \\
&\quad - \int dQ |\chi|^2 \I^{\mu\nu} g_{\mu\tau} \p_{\nu} \f{J^{\tau}}{|\chi|^2} {.}
\label{eq:final}
\end{align}
We used the identity
\begin{align}
C_{\tau\sigma\mu} = C_{\mu\sigma\tau} + \f{1}{2} \p_{\sigma} B_{\tau\mu} 
\end{align}
to combine the second and third terms of equation (\ref{eq:dEgeo:final}), after contraction with $\I^{\mu\nu}$, into the divergence of the vector field 
\begin{align}
- \f{1}{2} \I^{\nu\mu} \Big( C_{\mu\sigma\tau} |\chi|^2\Big) \I^{\sigma\tau} {,}
\end{align}
whose volume integral, converted into a surface integral via Gauss's theorem, generally vanishes for finite systems.

\section{One-dimensional example}

\begin{figure}[!h]
\centering\includegraphics[width=2.0in]{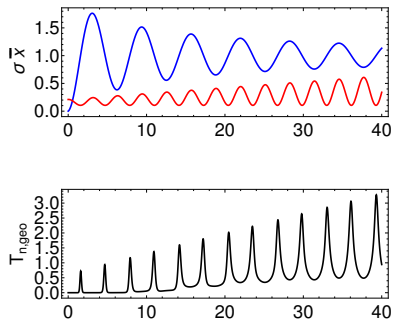}
\caption{Top panel: The mean position $\ox(t)$ and width $\sigma(t)$ of the gaussian nuclear wave packet are plotted versus time (a.u.) for the one-dimensional model system with $\eta=0.1$, $M=10$~a.u. and $\gamma=40$.  Bottom panel: The geometric part of the nuclear kinetic energy $T_{n,geo}$ versus time (a.u.) for the same parameters.}
\label{fig}
\end{figure}

We illustrate the identity (\ref{eq:final}) for an exactly solvable one-dimensional model with two electronic states.  The Schr\"odinger equation is
\begin{align}
i\p_t \Psi = -\f{1}{2} \I \p_x \p_x \Psi + \left( \begin{array}{cc} h_0 + h_3 & h_1 \\ h_1 & h_0 - h_3 \end{array} \right) \Psi {.} \label{eq:model}
\end{align}
Our strategy is to reverse engineer the functions $h_0(x,t)$, $h_1(x,t)$ and $h_3(x,t)$ such that the dynamics lead to a state 
\begin{align}
\Psi(x,t) = \chi(x,t) \left( \begin{array}{l} e^{i\alpha/2-i\varphi/2} \cos (\ot/2) \\ e^{i\alpha/2+i\varphi/2} \sin (\ot/2) \end{array} \right) 
\end{align}
with a gaussian nuclear density 
\begin{align}
|\chi(x,t)|^2 = \f{1}{\sqrt{\pi} \sigma(t)} \exp \bigg( -\f{[x-\ox(t)]^2}{\sigma^2(t)} \bigg)
\label{eq:chisq}
\end{align}
that undergoes damped oscillations determined by the functions
\begin{align}
\ox(t) &= 1 - \f{1}{1+\eta t} \cos t \nn \\
\sigma(t) &= \f{1}{3\sqrt{M}} \Big[ 1 + (1+\eta t) \cos^2 t \Big] {.}
\end{align}
Substituting into equation (\ref{eq:model}), leads to the following equations:
\begin{align}
\p_t \ln|\chi| &= -\f{1}{2} \I (\p_x \ln|\chi|) (\alpha_x - \cos\ot \varphi_x)
-\f{1}{4} \I (\alpha_{xx} - \cos\ot \varphi_{xx})
-\f{1}{4} \I \sin\ot \ot_x \varphi_x \nn \\
\ot_t &= -2h_1 \sin\varphi - \I \sin\ot (\p_x \ln|\chi|) \varphi_x - \f{1}{2} \I \sin\ot \varphi_{xx}
-\f{1}{2} \I \ot_x (\alpha_x + \cos\ot \varphi_x) \nn \\
\sin\ot \varphi_t &= 2(-h_1 \cos\ot \cos\varphi + h_3 \sin\ot) + \I (\p_x \ln|\chi|) \ot_x
-\f{1}{2} \I \sin\ot \alpha_x \varphi_x + \f{1}{2} \I \ot_{xx} \nn \\
\alpha_t - \cos\ot \varphi_t &= -2 ( h_0 + h_1 \sin\ot \cos \varphi + h_3 \cos\ot ) + \I \p_x^2 \ln|\chi| + \I (\p_x \ln|\chi|)^2  \nn \\
&\quad-\f{1}{4} \I \big[ \alpha_x^2 + \varphi_x^2 - 2 \cos\ot \alpha_x \varphi_x \big]
-\f{1}{4} \I \ot_x^2 
{,} \label{eq:implicit}
\end{align}
where subscripts denote partial differentiation, e.g.~$\alpha_x = \p_x \alpha$. The first equation is equivalent to the continuity equation $\p_t |\chi|^2 = -\p_x J$ with the nuclear current density
\begin{align}
J = \I |\chi|^2 A
\end{align}  
and vector potential 
\begin{align}
A = \f{1}{2} (\alpha_x - \cos\ot \varphi_x) {.}
\end{align}
With $|\chi(x,t)|^2$ given by equation (\ref{eq:chisq}), we have 
\begin{align} 
A(x,t) = -\I^{-1} \f{1}{|\chi(x,t)|^2} \int^x dx' \p_t |\chi(x',t)|^2 {.}
\end{align}
Defining $w = \cos\ot$, we choose a state $\Psi(x,t)$ with 
\begin{align}
w(x,t) &= \eta + (1-2\eta) \f{1+t}{1 + t + e^{\gamma(1+\eta t)(x-1)}} \nn \\
\varphi(x,t) &= -\eta - (1-2\eta) \f{1+3t}{1 + 3t + e^{\gamma(1+\eta t)(x-1)}} \nn \\
\alpha(x,t) &= \int^x dx' \big[ 2A(x',t) + w(x',t) \p_{x'} \varphi(x',t) \big] {.}
\end{align}
The Hamiltonian parameters that follow from equation (\ref{eq:implicit}) are
\begin{align}
h_0 &= -h_1 \sin\ot \cos\varphi - h_3 \cos\ot - \f{1}{2} \alpha_t + \f{1}{2} \cos\ot \varphi_t
+\f{1}{2} \I \p_x^2 \ln|\chi| +\f{1}{2} \I  (\p_x \ln|\chi|)^2 \nn \\
&\quad -\f{1}{8} \I \big[ \alpha_x^2 + \varphi_x^2 - 2 \cos\ot \alpha_x \varphi_x \big]
-\f{1}{8} \I \ot_x^2  \nn \\
h_1 &= \f{1}{\sin\varphi} \Big[ -\f{1}{2} \ot_t - \f{1}{2} \I \sin\ot (\p_x \ln|\chi|) \varphi_x
-\f{1}{4} \I \sin\ot \varphi_{xx} - \f{1}{4} \I \ot_x (\alpha_x + \cos\ot \varphi_x) \Big] \nn \\
h_3 &= \f{1}{\sin\ot} \Big[ h_1 \cos\ot \cos\varphi + \f{1}{2} \sin\ot \varphi_t 
-\f{1}{2} \I (\p_x \ln|\chi|) \ot_x + \f{1}{4} \I \sin\ot \alpha_x \varphi_x
-\f{1}{4} \I \ot_{xx} \Big] {.}
\end{align}
The identity in equation (\ref{eq:final}), adapted to the present model, is
\begin{align}
\f{dT_{n,geo}}{dt} 
&= -\int dx \I \,\Im \big< \Phi \big| \partial_{x} H^{BO} \big| \p_{x} \Phi \big> |\chi|^2 + \int dx \I A \big< \Phi \big| \partial_{x} H^{BO} \big| \Phi \big> - \f{1}{2} \int dx \I \I \p_x \big( C |\chi|^2 \big) \nn \\
&\quad - \int dx \I \I g \p_x A {.}
\label{eq:final:model}
\end{align}
The geometric quantities needed to evaluate the right-hand side are
\begin{align}
g &= \f{1}{4} \f{w_x^2}{1-w^2} + \f{1}{4} (1-w^2) \varphi_x^2 \nn \\
C &= -\f{1}{4} \f{1}{1-w^2} \Big[ - w (1-w^2)^2 \varphi_x^3 - 3 w w_x^2 \varphi_x + (1-w^2) (w_x \varphi_{xx} - w_{xx} \varphi_x) \Big] {,}
\end{align}
and for completeness we also record
\begin{align}
D &= -\f{1}{8} \f{1}{(1-w^2)^2} \Big[ 2 w w_x \big((1-w^2)^2 \varphi_x^2 + w_x^2\big) - (1-w^2) \big( 4 w (1-w^2) w_x \varphi_x^2 \nn \\
&\hspace{2.4cm}- 2 (1-w^2)^2 \varphi_x \varphi_{xx} - 2 w_x w_{xx} \big) \Big] {.}
\end{align}
Using the above formulas, we have numerically verified Eq.~(\ref{eq:final:model}) for $\eta=0.1$, $M=10$~a.u. and $\gamma=40$.  The functions $\ox(t)$ and $\sigma(t)$ defining the damped oscillations and the time dependence of $T_{n,geo}$ are shown in Fig.~\ref{fig}.

\section{Third-rank quantum geometric quantities \label{sec:CD}}

Here we briefly investigate some properties of the rank-3 quantities $C_{\mu\nu\tau}$ and $D_{\mu\nu\tau}$ that appeared in the last section.  Beginning with $D_{\mu\nu\tau}$, we can show that
\begin{align}
D_{\mu\nu\tau} 
&= -\Re \langle \p_{\mu} \Phi | \p_{\nu} \p_{\tau} \Phi \rangle - \f{1}{2} B_{\mu\nu} A_{\tau} - \f{1}{2} B_{\mu\tau} A_{\nu} + \f{1}{2} A_{\mu} \p_{\nu} A_{\tau} + \f{1}{2} A_{\mu} \p_{\tau} A_{\nu} {.} 
\label{eq:D:1}
\end{align}
Thus, $D_{\mu\nu\tau}$ is symmetric with respect to interchange of its second two indices.
To further simplify $D_{\mu\nu\tau}$, we derive the following identity:
\begin{align}
\langle \p_{\mu} \Phi | \p_{\nu} \p_{\tau} \Phi \rangle 
&= \p_{\nu} \langle \p_{\mu} \Phi | \p_{\tau} \Phi \rangle - \langle \p_{\mu} \p_{\nu} \Phi | \p_{\tau} \Phi \rangle \nn \\
&= \p_{\nu} \langle \p_{\mu} \Phi | \p_{\tau} \Phi \rangle - \p_{\mu} \langle \p_{\nu} \Phi | \p_{\tau} \Phi \rangle + \langle \p_{\nu} \Phi | \p_{\mu} \p_{\tau} \Phi \rangle \nn \\
&= \p_{\nu} \langle \p_{\mu}\Phi | \p_{\tau}\Phi \rangle - \p_{\mu} \langle \p_{\nu} \Phi | \p_{\tau} \Phi \rangle + \p_{\tau} \langle \p_{\nu} \Phi | \p_{\mu} \Phi \rangle -\langle \p_{\nu} \p_{\tau} \Phi | \p_{\mu} \Phi \rangle \nn \\
&= -\langle \p_{\mu} \Phi | \p_{\nu} \p_{\tau} \Phi \rangle^* + \p_{\tau} \langle \p_{\nu} \Phi | \p_{\mu} \Phi \rangle + \p_{\nu} \langle \p_{\mu}\Phi | \p_{\tau}\Phi \rangle - \p_{\mu} \langle \p_{\nu} \Phi | \p_{\tau} \Phi \rangle {.}
\label{eq:identity:2}
\end{align}
Taking the real part, we find
\begin{align}
2\Re \langle \p_{\mu} \Phi | \p_{\nu} \p_{\tau} \Phi \rangle = \p_{\tau} (g_{\mu\nu}+A_{\mu}A_{\nu}) + \p_{\nu} (g_{\mu\tau}+A_{\mu}A_{\tau}) - \p_{\mu} (g_{\nu\tau}+A_{\nu}A_{\tau}) {.}
\end{align}
With this identity and equation (\ref{eq:D:1}), we obtain 
\begin{align}
D_{\mu\nu\tau} &= -\f{1}{2} \p_{\tau} g_{\mu\nu} - \f{1}{2} \p_{\nu} g_{\mu\tau} + \f{1}{2} \p_{\mu} g_{\nu\tau}  {.}
\label{eq:D:2}
\end{align}
Thus, $D_{\mu\nu\tau}$ has been expressed in terms of $g_{\mu\nu}$.  In fact, we have 
\begin{align}
D_{\mu\nu\tau} = -\Gamma_{\mu\nu\tau} {,}
\end{align}
where $\Gamma_{\mu\nu\tau}$ is the Christoffel symbol of the first kind in classical Riemannian geometry.

Turning to $C_{\mu\nu\tau}$, we find the expression
\begin{align}
C_{\mu\nu\tau} &= \Im \langle \p_{\mu} \Phi | \p_{\nu} \p_{\tau} \Phi \rangle - A_{\mu} g_{\nu\tau} - A_{\nu} g_{\mu\tau} - A_{\tau} g_{\mu\nu} - A_{\mu} A_{\nu} A_{\tau} {.}
\label{eq:C:2}
\end{align}
This is also symmetric with respect to the interchange of the second two indices.
However, due to the presence of the irreducible third-order quantity $\Im \langle \p_{\mu} \Phi | \p_{\nu} \p_{\tau} \Phi \rangle$, $C_{\mu\nu\tau}$ cannot be expressed in terms of lower-order geometric quantities and their derivatives.  

\section{Conclusion}

We derived an identity for the rate that energy is transferred to $T_{n,geo}$, the so-called geometric part of the nuclear kinetic energy.  This is the part that derives from the gradient with respect to a nuclear coordinate acting on the parametric dependence of an electronic wave function, the latter arising from the factorization of the full wave function.  Our identity complements the Ehrenfest-like identity previously derived \cite{li2019arxiv} for $dT_{n,marg}/dt$, the ``marginal'' part of the nuclear kinetic energy.

Ehrenfest identities for the expectation values of position and momentum resemble Newton's laws.  Similarly, the Ehrenfest-like identity for $dT_{n,marg}/dt$ has a suggestive force-times-velocity form, paralleling the classical formula for the rate of work done by a force, and therefore appears to lend itself to a classical interpretation of the nuclear motion.  Indeed, the terms that appear in the force operator have a close resemblance to the corresponding terms in the force on classical nuclei \cite{agostini2014,abedi2014}.  

It is not yet clear if identity (\ref{eq:final}) for $dT_{n,geo}/dt$ has a simple classical interpretation.  We have not been able to put it in a force-times-velocity form.  In the course of evaluating $dT_{n,geo}/dt$, we derived the equation of motion for the quantum metric $g_{\mu\nu}$.  This equation involves a new quantity, the rank-3 geometric quantity $C_{\mu\nu\tau}$, which appears to be a purely quantum object.  

Putting together the identities for $dT_{n,marg}/dt$ and $dT_{n,geo}/dt$ allow us to calculate the total rate of change of the nuclear kinetic energy.  It is hoped that these identities and the insights derived from them will help scientists control energy transfer in quantum systems.

\begin{acknowledgments}

This work received funding from the European Research Council (ERC) under the European Union's Horizon 2020 research and innovation programme (grant agreement No.~ERC-2017-AdG-788890).

\end{acknowledgments}

\bibliography{Egeo}

\begin{thebibliography}{18}%
\makeatletter
\providecommand \@ifxundefined [1]{%
 \@ifx{#1\undefined}
}%
\providecommand \@ifnum [1]{%
 \ifnum #1\expandafter \@firstoftwo
 \else \expandafter \@secondoftwo
 \fi
}%
\providecommand \@ifx [1]{%
 \ifx #1\expandafter \@firstoftwo
 \else \expandafter \@secondoftwo
 \fi
}%
\providecommand \natexlab [1]{#1}%
\providecommand \enquote  [1]{``#1''}%
\providecommand \bibnamefont  [1]{#1}%
\providecommand \bibfnamefont [1]{#1}%
\providecommand \citenamefont [1]{#1}%
\providecommand \href@noop [0]{\@secondoftwo}%
\providecommand \href [0]{\begingroup \@sanitize@url \@href}%
\providecommand \@href[1]{\@@startlink{#1}\@@href}%
\providecommand \@@href[1]{\endgroup#1\@@endlink}%
\providecommand \@sanitize@url [0]{\catcode `\\12\catcode `\$12\catcode
  `\&12\catcode `\#12\catcode `\^12\catcode `\_12\catcode `\%12\relax}%
\providecommand \@@startlink[1]{}%
\providecommand \@@endlink[0]{}%
\providecommand \url  [0]{\begingroup\@sanitize@url \@url }%
\providecommand \@url [1]{\endgroup\@href {#1}{\urlprefix }}%
\providecommand \urlprefix  [0]{URL }%
\providecommand \Eprint [0]{\href }%
\providecommand \doibase [0]{http://dx.doi.org/}%
\providecommand \selectlanguage [0]{\@gobble}%
\providecommand \bibinfo  [0]{\@secondoftwo}%
\providecommand \bibfield  [0]{\@secondoftwo}%
\providecommand \translation [1]{[#1]}%
\providecommand \BibitemOpen [0]{}%
\providecommand \bibitemStop [0]{}%
\providecommand \bibitemNoStop [0]{.\EOS\space}%
\providecommand \EOS [0]{\spacefactor3000\relax}%
\providecommand \BibitemShut  [1]{\csname bibitem#1\endcsname}%
\let\auto@bib@innerbib\@empty
\bibitem [{\citenamefont {Gidopoulos}\ and\ \citenamefont
  {Gross}(2014)}]{gidopoulos2014}%
  \BibitemOpen
  \bibfield  {author} {\bibinfo {author} {\bibfnamefont {N.~I.}\ \bibnamefont
  {Gidopoulos}}\ and\ \bibinfo {author} {\bibfnamefont {E.~K.~U.}\ \bibnamefont
  {Gross}},\ }\href@noop {} {\bibfield  {journal} {\bibinfo  {journal} {Phil.
  Trans. Roy. Soc. A}\ }\textbf {\bibinfo {volume} {372}},\ \bibinfo {pages}
  {20130059} (\bibinfo {year} {2014})}\BibitemShut {NoStop}%
\bibitem [{\citenamefont {Abedi}\ \emph {et~al.}(2010)\citenamefont {Abedi},
  \citenamefont {Maitra},\ and\ \citenamefont {Gross}}]{abedi2010}%
  \BibitemOpen
  \bibfield  {author} {\bibinfo {author} {\bibfnamefont {A.}~\bibnamefont
  {Abedi}}, \bibinfo {author} {\bibfnamefont {N.~T.}\ \bibnamefont {Maitra}}, \
  and\ \bibinfo {author} {\bibfnamefont {E.~K.~U.}\ \bibnamefont {Gross}},\
  }\href@noop {} {\bibfield  {journal} {\bibinfo  {journal} {Phys. Rev. Lett.}\
  }\textbf {\bibinfo {volume} {105}},\ \bibinfo {pages} {123002} (\bibinfo
  {year} {2010})}\BibitemShut {NoStop}%
\bibitem [{\citenamefont {Hunter}(1975)}]{hunter1975}%
  \BibitemOpen
  \bibfield  {author} {\bibinfo {author} {\bibfnamefont {G.}~\bibnamefont
  {Hunter}},\ }\href@noop {} {\bibfield  {journal} {\bibinfo  {journal} {Int.
  J. Quantum Chem.}\ }\textbf {\bibinfo {volume} {9}},\ \bibinfo {pages} {237}
  (\bibinfo {year} {1975})}\BibitemShut {NoStop}%
\bibitem [{\citenamefont {Li}\ \emph {et~al.}(2019)\citenamefont {Li},
  \citenamefont {Requist},\ and\ \citenamefont {Gross}}]{li2019arxiv}%
  \BibitemOpen
  \bibfield  {author} {\bibinfo {author} {\bibfnamefont {C.}~\bibnamefont
  {Li}}, \bibinfo {author} {\bibfnamefont {R.}~\bibnamefont {Requist}}, \ and\
  \bibinfo {author} {\bibfnamefont {E.~K.~U.}\ \bibnamefont {Gross}},\
  }\href@noop {} {\enquote {\bibinfo {title} {Energy, momentum and angular
  momentum transfer between electrons and nuclei},}\ }\bibinfo {howpublished}
  {arxiv:1908.04077v1} (\bibinfo {year} {2019})\BibitemShut {NoStop}%
\bibitem [{\citenamefont {Abedi}\ \emph {et~al.}(2012)\citenamefont {Abedi},
  \citenamefont {Maitra},\ and\ \citenamefont {Gross}}]{abedi2012}%
  \BibitemOpen
  \bibfield  {author} {\bibinfo {author} {\bibfnamefont {A.}~\bibnamefont
  {Abedi}}, \bibinfo {author} {\bibfnamefont {N.~T.}\ \bibnamefont {Maitra}}, \
  and\ \bibinfo {author} {\bibfnamefont {E.~K.~U.}\ \bibnamefont {Gross}},\
  }\href@noop {} {\bibfield  {journal} {\bibinfo  {journal} {J. Chem. Phys.}\
  }\textbf {\bibinfo {volume} {137}},\ \bibinfo {pages} {22A530} (\bibinfo
  {year} {2012})}\BibitemShut {NoStop}%
\bibitem [{\citenamefont {Agostini}\ \emph {et~al.}(2015)\citenamefont
  {Agostini}, \citenamefont {Abedi}, \citenamefont {Suzuki}, \citenamefont
  {Min}, \citenamefont {Maitra},\ and\ \citenamefont {Gross}}]{agostini2015}%
  \BibitemOpen
  \bibfield  {author} {\bibinfo {author} {\bibfnamefont {F.}~\bibnamefont
  {Agostini}}, \bibinfo {author} {\bibfnamefont {A.}~\bibnamefont {Abedi}},
  \bibinfo {author} {\bibfnamefont {Y.}~\bibnamefont {Suzuki}}, \bibinfo
  {author} {\bibfnamefont {S.~K.}\ \bibnamefont {Min}}, \bibinfo {author}
  {\bibfnamefont {N.~T.}\ \bibnamefont {Maitra}}, \ and\ \bibinfo {author}
  {\bibfnamefont {E.~K.~U.}\ \bibnamefont {Gross}},\ }\href@noop {} {\bibfield
  {journal} {\bibinfo  {journal} {J. Chem. Phys.}\ }\textbf {\bibinfo {volume}
  {142}},\ \bibinfo {pages} {084303} (\bibinfo {year} {2015})}\BibitemShut
  {NoStop}%
\bibitem [{\citenamefont {Requist}\ \emph {et~al.}(2016)\citenamefont
  {Requist}, \citenamefont {Tandetzky},\ and\ \citenamefont
  {Gross}}]{requist2016a}%
  \BibitemOpen
  \bibfield  {author} {\bibinfo {author} {\bibfnamefont {R.}~\bibnamefont
  {Requist}}, \bibinfo {author} {\bibfnamefont {F.}~\bibnamefont {Tandetzky}},
  \ and\ \bibinfo {author} {\bibfnamefont {E.~K.~U.}\ \bibnamefont {Gross}},\
  }\href@noop {} {\bibfield  {journal} {\bibinfo  {journal} {Phys. Rev. A}\
  }\textbf {\bibinfo {volume} {93}},\ \bibinfo {pages} {042108} (\bibinfo
  {year} {2016})}\BibitemShut {NoStop}%
\bibitem [{\citenamefont {Requist}\ and\ \citenamefont
  {Gross}(2016)}]{requist2016b}%
  \BibitemOpen
  \bibfield  {author} {\bibinfo {author} {\bibfnamefont {R.}~\bibnamefont
  {Requist}}\ and\ \bibinfo {author} {\bibfnamefont {E.~K.~U.}\ \bibnamefont
  {Gross}},\ }\href@noop {} {\bibfield  {journal} {\bibinfo  {journal} {Phys.
  Rev. Lett.}\ }\textbf {\bibinfo {volume} {117}},\ \bibinfo {pages} {193001}
  (\bibinfo {year} {2016})}\BibitemShut {NoStop}%
\bibitem [{\citenamefont {Provost}\ and\ \citenamefont
  {Vallee}(1980)}]{provost1980}%
  \BibitemOpen
  \bibfield  {author} {\bibinfo {author} {\bibfnamefont {J.~P.}\ \bibnamefont
  {Provost}}\ and\ \bibinfo {author} {\bibfnamefont {G.}~\bibnamefont
  {Vallee}},\ }\href@noop {} {\bibfield  {journal} {\bibinfo  {journal}
  {Commun. Math. Phys.}\ }\textbf {\bibinfo {volume} {76}},\ \bibinfo {pages}
  {289} (\bibinfo {year} {1980})}\BibitemShut {NoStop}%
\bibitem [{\citenamefont {Watson}(1968)}]{watson1968}%
  \BibitemOpen
  \bibfield  {author} {\bibinfo {author} {\bibfnamefont {J.~K.~G.}\
  \bibnamefont {Watson}},\ }\href@noop {} {\bibfield  {journal} {\bibinfo
  {journal} {Molec. Phys.}\ }\textbf {\bibinfo {volume} {15}},\ \bibinfo
  {pages} {479} (\bibinfo {year} {1968})}\BibitemShut {NoStop}%
\bibitem [{\citenamefont {Berry}(1989)}]{berry1989}%
  \BibitemOpen
  \bibfield  {author} {\bibinfo {author} {\bibfnamefont {M.~V.}\ \bibnamefont
  {Berry}},\ }\enquote {\bibinfo {title} {The quantum phase, five years
  after},}\ \ (\bibinfo {year} {1989})\ pp.\ \bibinfo {pages} {7--28},\
  \bibinfo {note} {in ref.~\cite{shapere1989}}\BibitemShut {NoStop}%
\bibitem [{\citenamefont {Shapere}\ and\ \citenamefont
  {Wilczek}(1989)}]{shapere1989}%
  \BibitemOpen
  \bibinfo {editor} {\bibfnamefont {A.}~\bibnamefont {Shapere}}\ and\ \bibinfo
  {editor} {\bibfnamefont {F.}~\bibnamefont {Wilczek}},\ eds.,\ \href@noop {}
  {\emph {\bibinfo {title} {Geometric phases in physics}}}\ (\bibinfo
  {publisher} {World Scientific, Singapore},\ \bibinfo {year}
  {1989})\BibitemShut {NoStop}%
\bibitem [{\citenamefont {Berry}\ and\ \citenamefont {Lim}(1990)}]{berry1990}%
  \BibitemOpen
  \bibfield  {author} {\bibinfo {author} {\bibfnamefont {M.~V.}\ \bibnamefont
  {Berry}}\ and\ \bibinfo {author} {\bibfnamefont {R.}~\bibnamefont {Lim}},\
  }\href@noop {} {\bibfield  {journal} {\bibinfo  {journal} {J. Phys. A: Math.
  Gen}\ }\textbf {\bibinfo {volume} {23}},\ \bibinfo {pages} {L655} (\bibinfo
  {year} {1990})}\BibitemShut {NoStop}%
\bibitem [{\citenamefont {Berry}\ and\ \citenamefont
  {Robbins}(1993)}]{berry1993}%
  \BibitemOpen
  \bibfield  {author} {\bibinfo {author} {\bibfnamefont {M.~V.}\ \bibnamefont
  {Berry}}\ and\ \bibinfo {author} {\bibfnamefont {J.~M.}\ \bibnamefont
  {Robbins}},\ }\href@noop {} {\bibfield  {journal} {\bibinfo  {journal} {Proc.
  R. Soc. Lond. A}\ }\textbf {\bibinfo {volume} {442}},\ \bibinfo {pages} {641}
  (\bibinfo {year} {1993})}\BibitemShut {NoStop}%
\bibitem [{\citenamefont {Sutcliffe}(2000)}]{sutcliffe2000}%
  \BibitemOpen
  \bibfield  {author} {\bibinfo {author} {\bibfnamefont {B.~T.}\ \bibnamefont
  {Sutcliffe}},\ }\href@noop {} {\bibfield  {journal} {\bibinfo  {journal}
  {Adv. Chem. Phys.}\ }\textbf {\bibinfo {volume} {114}},\ \bibinfo {pages}
  {97} (\bibinfo {year} {2000})}\BibitemShut {NoStop}%
\bibitem [{\citenamefont {Requist}\ \emph {et~al.}(2019)\citenamefont
  {Requist}, \citenamefont {Proetto},\ and\ \citenamefont
  {Gross}}]{requist2019}%
  \BibitemOpen
  \bibfield  {author} {\bibinfo {author} {\bibfnamefont {R.}~\bibnamefont
  {Requist}}, \bibinfo {author} {\bibfnamefont {C.~R.}\ \bibnamefont
  {Proetto}}, \ and\ \bibinfo {author} {\bibfnamefont {E.~K.~U.}\ \bibnamefont
  {Gross}},\ }\href@noop {} {\bibfield  {journal} {\bibinfo  {journal} {Phys.
  Rev. B}\ }\textbf {\bibinfo {volume} {99}},\ \bibinfo {pages} {165136}
  (\bibinfo {year} {2019})}\BibitemShut {NoStop}%
\bibitem [{\citenamefont {Agostini}\ \emph {et~al.}(2014)\citenamefont
  {Agostini}, \citenamefont {Abedi},\ and\ \citenamefont
  {Gross}}]{agostini2014}%
  \BibitemOpen
  \bibfield  {author} {\bibinfo {author} {\bibfnamefont {F.}~\bibnamefont
  {Agostini}}, \bibinfo {author} {\bibfnamefont {A.}~\bibnamefont {Abedi}}, \
  and\ \bibinfo {author} {\bibfnamefont {E.~K.~U.}\ \bibnamefont {Gross}},\
  }\href@noop {} {\bibfield  {journal} {\bibinfo  {journal} {J. Chem. Phys.}\
  }\textbf {\bibinfo {volume} {141}},\ \bibinfo {pages} {214101} (\bibinfo
  {year} {2014})}\BibitemShut {NoStop}%
\bibitem [{\citenamefont {Abedi}\ \emph {et~al.}(2014)\citenamefont {Abedi},
  \citenamefont {Agostini},\ and\ \citenamefont {Gross}}]{abedi2014}%
  \BibitemOpen
  \bibfield  {author} {\bibinfo {author} {\bibfnamefont {A.}~\bibnamefont
  {Abedi}}, \bibinfo {author} {\bibfnamefont {F.}~\bibnamefont {Agostini}}, \
  and\ \bibinfo {author} {\bibfnamefont {E.~K.~U.}\ \bibnamefont {Gross}},\
  }\href@noop {} {\bibfield  {journal} {\bibinfo  {journal} {Europhys. Lett.}\
  }\textbf {\bibinfo {volume} {106}},\ \bibinfo {pages} {33001} (\bibinfo
  {year} {2014})}\BibitemShut {NoStop}%
\end{thebibliography}%

\end{document}